\documentclass{article}

\usepackage[utf8]{inputenc} 
\usepackage[T1]{fontenc}    
\usepackage{hyperref}       
\usepackage{url}            
\usepackage{booktabs}       
\usepackage{amsfonts}       
\usepackage{nicefrac}       
\usepackage{microtype}      
\usepackage{amsmath}
\usepackage{amsfonts}
\usepackage{listofitems}
\usepackage{graphicx}%
\usepackage{float}

\newtheorem{theorem}{Theorem}[section]
\newtheorem{lemma}[theorem]{Lemma}

\newtheorem{defn}[theorem]{Definition}

\newcommand{\be}{\begin{equation}}
\newcommand{\ee}{\end{equation}}

\newcommand{\myX}{\boldsymbol{X}}
\newcommand{\myx}{\boldsymbol{x}}

\newcommand{\myE}{\mathbb{E}}
\newcommand{\myV}{\mathbb{V}}

\newcommand{\mytheta}{\boldsymbol{\theta}}
\newcommand{\myregtheta}{\mytheta^*}

\newcommand{\mymletheta}{\hat{\mytheta}}

\newcommand{\myTheta}{\boldsymbol{\Theta}}

\newcommand{\data}{\myx_n}

\newcommand{\var}{x}

\newcommand{\bVar}{\myX}
\newcommand{\Var}{\myX}

\newcommand{\newVar}{\Var'} 

\newcommand{\model}[1]{%
 \readlist*\myargs{#1}%
  g(\myargs[1] | \myargs[2])}

\DeclareMathOperator*{\argmin}{argmin}

\newenvironment{proof}[1][Proof]{\begin{trivlist}
\item[\hskip \labelsep {\bfseries #1}]}{\end{trivlist}}

\newenvironment{remark}[1][Remark]{\begin{trivlist}
\item[\hskip \labelsep {\bfseries #1}]}{\end{trivlist}}

\newcommand{\qed}{\nobreak \ifvmode \relax \else
      \ifdim\lastskip<1.5em \hskip-\lastskip
      \hskip1.5em plus0em minus0.5em \fi \nobreak
      \vrule height0.75em width0.5em depth0.25em\fi}

\usepackage[framemethod=TikZ]{mdframed}

\newmdenv[%
    backgroundcolor=red!8,
    linecolor=red,
    outerlinewidth=1pt,
    roundcorner=5mm,
    skipabove=\baselineskip,
    skipbelow=\baselineskip,
]{redbox}

\begin{document}

\title{\bf Information-Corrected Estimation: A Generalization Error Reducing Parameter Estimation Method}

\author{Matthew F. Dixon\thanks{Department of Applied Mathematics, Illinois Institute of Technology, Chicago, IL 60616-3793, USA; matthew.dixon@iit.edu.}\\
     \\
    and\\
    \\
    Tyler Ward\thanks{Department of Financial Engineering, NYU Tandon School of Engineering, New York, NY 11201, USA. E-mail:  tw623@nyu.edu.}\\
    }
  \maketitle

\begin{abstract}
Modern computational models in supervised machine learning are often highly parameterized universal approximators. As such, the value of the parameters is unimportant, and only the out of sample performance is considered. On the other hand, much of the literature on model estimation assumes that the parameters themselves have intrinsic value, and thus is concerned with bias and variance of parameter estimates, which may not have any simple relationship to out of sample model performance. Therefore, within supervised machine learning, heavy use is made of ridge regression (i.e., L2 regularization), which requires the the estimation of hyperparameters and can be rendered ineffective by certain model parameterizations. We introduce an objective function which we refer to as Information-Corrected Estimation (ICE) that reduces KL divergence based generalization error for supervised machine learning. ICE attempts to directly maximize a corrected likelihood function as an estimator of the KL divergence.   Such an approach is proven, theoretically, to be effective for a wide class of models, with only mild regularity restrictions. Under finite sample sizes, this corrected estimation procedure is shown experimentally to lead to significant reduction in generalization error compared to maximum likelihood estimation and L2 regularization.

\end{abstract}

\section{Introduction}
\label{intro}

Kullback and Leibler \cite{Kullback1951} showed that minimizing a divergence $ \rho_{KL}(f, g_{\mytheta}) $ between the truth, $f$, and a parametric model density, $g_{\mytheta}$, is necessary and sufficient for making accurate predictions about data using the model defined by $ \mytheta $. Recent work \cite{BERKNASH} on Berk--Nash equilibria has shown the central role that KL divergence plays in game theoretic choice models such as multi-armed bandits and stochastic multi-party games. KL divergence thus plays a leading role in machine learning and  neuroscience, with several inferential approaches developed in the information theory literature.  Such approaches for minimizing KL divergence employ a range of methods, including data partitioning, Bayesian indirect inference and M-estimation \cite{pmlr-v84-jiang18a,Nguyen:2010:EDF:1921943.1921980,PrezCruz2008KullbackLeiblerDE}.  These approaches are quite distinct from the standard penalized loss minimization framework and, as such, are non-trivial to combine with supervised learning methods such as neural networks.

It is well known that maximum likelihood estimation (MLE) introduces an asymptotic bias in the KL divergence minimizer which is problematic for both model estimation and model selection. For many models, where the parameters $ \mytheta $ are themselves important, this may be investigated as parameter bias and parameter variance. However, for models common in modern machine learning, the parameters themselves do not have any easily interpreted meaning. For these models, the parameters themselves are irrelevant and only the accuracy (in terms of KL divergence) of the model predictions matter. Within the information theory literature, this has often been referred to simply as bias (e.g., $ b(G) $ from \cite{Konishi1996}). To distinguish it from parameter bias, one might refer to it as ``prediction bias'' or ``generalization error''. Generalization error is the more common terminology (see, for example, Equation 1.1.6 \cite{Roelofs:EECS-2019-102}) and will be used here. 

Before the widespread use of machine learning, most models had interpretable parameters, and thus there is a large literature focused on reducing parameter bias.  For instance, the jackknife~\cite{jackknife} (leave-one-out cross-validation) estimator is an early example. More relevant to this paper is the approach of Firth \cite{Firth93} and later Kosmidis \cite{KosThesis,kosmidis2010}. More recently, Pagui, Salvan, and Sartori \cite{10.1093/biomet/asx046} proposed a parameter bias reducing estimation methodology. An extensive review of the literature around this point can be found in \cite{Kosmidis_2014}. Unfortunately, these approaches do not consider the impact on KL divergence-based generalization error and thus are not applicable to the field of machine learning where the parameters themselves are devoid of meaning. Heskes \cite{Heskes98} shows that classifiers do have a notion of bias-variance decomposition for generalization error, but it is not computable from parameter bias and parameter variance. Therefore, parameter bias reducing formulations are not useful within machine learning unless it can be shown that they also reduce generalization error. 

In fact, to seat the approach taken in this paper to generalization error, we recall much earlier and seminal work at the intersection of statistics and information theory. Akaike \cite{Akaike1973}, and later Takeuchi \cite{Takeuchi1976}, proposed information criteria (AIC and TIC, respectively) for model selection designed explicitly to reduce generalization error. Konishi and Kitagawa \cite{Konishi1996} extended the approach of Takeuchi to cases where MLE was not used to fit the underlying model, but still restricted themselves to the question of model selection. Stone \cite{Stone1977} proved that Akaike's Information Criterion (AIC) is asymptotically equivalent to jackknifing when the estimator is finite. Takeuchi himself showed that TIC is an extension of AIC with fewer restrictions, and thus it too is equivalent to jackknifing whenever AIC would be valid. 

For highly parameterized models, as are common in machine learning, model selection such as this is of limited utility. The parameter count may necessarily be very large, and thus none of the models fit using MLE may be acceptable. Then, merely choosing among them is unlikely to produce acceptable results. Within this field, typically $ L_{2} $ or similar regularization is used to reduce generalization error. See Section 11.5.2 \cite{ESL-Hastie}, for a typical example. For a more recent innovation, refer to \cite{Ross_Doshi-Velez_2018}. Note that regularization schemes such as this often increase parameter bias while decreasing generalization error. Golub, Heath, and Wahba \cite{Golub1979} showed that $ L_{2} $ regularization is asymptotically equivalent to cross-validation for linear models, subject to certain assumptions. For nonlinear models, it has long been known that $ L_{2} $ regularization is not always valid, and it is trivial to construct example models\footnote{See Section \ref{gaussian_example} for one such example.}
 where this approach is always harmful in expectation. 

Therefore, it is important to develop a method to reduce generalization error in model estimation analogous to the way that $ L_{2} $ regularization would commonly be used for a highly parameterized model, but having applicability for a wider family of models, especially those for which $ L_{2} $ regularization is not applicable. It is not the goal of this paper to perform a wide survey of generalization error reducing approaches, but we will rather propose an additional approach, investigate its properties, and show that it has superior performance when compared against $ L_{2} $ regularization, which is currently the dominant generalization error reducing estimation procedure within the field of machine learning. 

To this end, this paper introduces a generalization error reducing estimation approach referred to as Information Corrected Estimation (ICE). This estimator is proven to have a generalization error of only $ O(n^{-\frac{3}{2}}) $ instead of $ O(n^{-1}) $ as is the case for MLE, and is shown to be valid within a neighborhood around the MLE parameter estimate. Optimizing over this ICE objective function instead of the negative log likelihood thus produces parameters with superior out of sample performance. 

Takeuchi's TIC and Firth's approach have never seen widespread use due to the computational and numerical issues that arise from the computation of this adjustment~\cite{YANAGIHARA20061965}, and the ICE estimator in its raw form would have similar problems. Therefore, this paper also proposes an efficient approximation of this correction term, and shows through numerical experiments that the approximation is effective at improving model performance across a range of models.


\section{Preliminaries}\label{sect:prelim}

Let us assume that we have data $\data:=\{\var_1,\dots, \var_n\}$ generated from an unknown joint density function $f(\myx)$ of $\Var_n:= \{X_1,\dots, X_n\}$. Where necessary, we define $ \newVar_n $ to denote a second sample drawn from $ f(\myx) $,  independent of $ \Var_n $, and  $ \data' $ is the observed realization of $ \newVar_n $. We consider a model $\mathcal{M}_p$ given by a parametric family of densities $\mathcal{M}_p:=\{\model{\cdot, \mytheta}~|~\mytheta\in\myTheta\subseteq\mathbb{R}^p\}$, for some compact Euclidean parameter space $\myTheta$, which is misspecified and hence excludes the truth $f$. Henceforth, the distribution over $ \var $ identified by $ \mytheta $ may be referred to as $ g_{\mytheta}(\myx) := g(\myx | \mytheta) $ where it is notationally convenient to do so. 

Suppose that $ \mytheta_0 $ is the quasi-true parameter of model $\mathcal{M}$, and $ \mymletheta(\Var_n) $ is the random variable representing the MLE of $ \mytheta_0 $ fit on a dataset, $ \data $. The negative log-likelihood of $ \Var_n $ under the distribution $ g_{\mytheta} $ is 

\be
\label{mleEquation}
-\ell(\mytheta, \Var_n) := -\frac{1}{n} \sum_{i=1}^{n} \log g_{\mytheta}(\myx_{i}),
\ee
where $ -\ell(\mytheta, \Var_n) $ is written including a $ \frac{1}{n} $ to make the expectation of this quantity $ O(1) $ and asymptotically independent of $ n $. Similarly, the minus sign is incorporated because $ -\ell(\mytheta, \Var_n) $ is a strictly non-negative quantity if $ g_{\mytheta}(\myx_{i}) $ is a probability. The MLE, $ \mymletheta(\Var_n)$, minimizes the negative log likelihood of the data set with respect to the model:

\be
\mymletheta(\data) := \argmin_{\mytheta}[-\ell(\mytheta, \data)].
\ee
The expectation of $ -\ell(\mytheta, \Var_n)$ is the cross entropy between $ f $ and $ g_{\mytheta} $:
\be
-\mathcal{L}(\mytheta) := \myE_{\Var_n}[-\ell(\mytheta, \Var_n)].
\ee

Here, the expectation is a function only of $ \mytheta $ and of the distribution $ f $ that generated the data $ \Var_n $. As a function of the distribution $ f $, this value is $ O(1) $, but could be  large for poorly conditioned $ f $. The quasi-true parameter $ \mytheta_0 $ is 

\be
\mytheta_0 := \argmin_{\mytheta}[-\mathcal{L}(\mytheta)].
\ee

\subsection*{Generalization Error in KL Divergence Based Loss Functions}

Kullback and Leibler \cite{Kullback1951} viewed ``information'' as discriminating the sample data drawn from one distribution against another, and defined the KL-divergence $ \rho_{KL} $ between distributions in terms of the ability to make predictions about one by knowing the other. Here, 
\begin{equation}
\rho_{KL}(f, g_{\mytheta}) = \int \log[\frac{f(x)}{g_{\mytheta}(x)}] f(x) dx.
\end{equation}
This value is in general unknowable, but given a sample $ \Var_n $ from $ f $, $ -\ell(\mytheta, \Var_n) $ will converge asymptotically to $ \rho_{KL}(f, g_{\mytheta}) $ plus an additive constant that depends only on $ f $. The convergence relies on White's regularity conditions \cite{White1982}.


A well known result by Stone \cite{Stone1977} shows that the MLE is a biased estimator of the minimum KL-divergence:

\be
\myE_{\Var_n}[-\ell(\mymletheta(\Var_n), \Var_n)] < \myE_{\Var_n}[-\ell(\mytheta_0, \Var_n)],
\ee
because it is evaluated on the data $ \Var_n $ which was used to fit $ \mymletheta $. Cross-validation was developed as a model selection technique to select a model from a group that actually minimizes $ \myE_{\Var_n}[\rho_{KL}(g_{\mytheta_0}, g_{\mymletheta(\Var_n)})] $ and not merely $ \myE_{\Var_n}[-\ell(\mymletheta(\Var_n), \Var_n)] $ in the limit of large $n$. Takeuchi \cite{Takeuchi1976} and Akaike \cite{Akaike1973} explicitly modeled this bias (generalization error) of an estimation procedure $ \mytheta(\Var_n) $ as 

\be
b := \myE_{\Var_n}\left[\ell(\mytheta(\Var_n), \Var_n) - \myE_{\Var'_n}[\ell(\mytheta(\Var_n), \Var'_n)]\right].
\ee

Our goal is to obtain an estimate, $ b^*$, of the generalization error $ b $ without using the MLE. We will then add this term to the objective function to develop the estimator $ \myregtheta(\Var_n) $ so as to cancel the lower order terms of the generalization error. This estimator will then minimize $ \myE_{\Var_n}[\rho_{KL}(g_{\mytheta_0}, g_{\myregtheta(\Var_n)})] $ more effectively than MLE, and potentially would in turn produce improved predictions from the model fitted over finite training sets. 

\begin{remark}
We note that under MLE, $ b = O(\frac{1}{n}) $ \cite{Takeuchi1976}. Equivalently, one could say that a particular realization of the generalization error $ \ell(\mytheta(\Var_n), \Var_n) - \myE_{\Var'_n}[\ell(\mytheta(\Var_n), \Var'_n)] $ is itself $ O_{p}(\frac{1}{n})  $. Here, $ O_{p}(\frac{1}{n}) $ is used to indicate that the quantity is a random variable with finite variance, whose mean is $ O(\frac{1}{n}) $.
\end{remark}

\section{Information Corrected Estimation (ICE)}
\label{iceSection}

We propose the following penalized likelihood function:

\begin{defn}[ICE Objective]
\be 
\label{eq:ice}
-\ell^{*}(\mytheta) = -\ell(\mytheta) + \frac{1}{n}tr(I_{\mytheta}J_{\mytheta}^{-1}),
\ee
where $ J_{\mytheta} $ is the negative expected Hessian 
\be
J_{\mytheta} := - \myE_{\Var}[\partial^2_{\mytheta}\log \model{\Var, \mytheta}]=-\int f(\var) \partial^2_{\mytheta} \log \model{\var,\mytheta} d\var,
\ee
and $ I_{\mytheta} $ is the Fisher Information matrix 
\be
I_{\mytheta} :=\myE_{\Var}[\partial_{\mytheta}\log \model{\Var, \mytheta} \partial_{\mytheta^T}\log \model{\Var, \mytheta}].
\ee
with $ \hat{I}_{\mytheta} $, $ \hat{J}_{\mytheta} $ being their estimates over the data. 
 \\
Let $\myregtheta$ denote the minimizer of \eqref{eq:ice}.
\end{defn}

The trace term in Equation (\ref{eq:ice}) will be familiar from Takeuchi \cite{Takeuchi1976}. However, Takeuchi showed only that this was the leading order of the bias for the MLE estimate $ \mymletheta $, and therefore the proof found there is not sufficient to justify a new estimator that will itself be the target of optimization, and is required to be valid away from $ \mymletheta $. As in Takeuchi, because $ I $ and $ J $ are unknowable, we will substitute their approximations computed from the training data, $ \hat{I_{\mytheta}} $ and $ \hat{J_{\mytheta}} $ during the actual computation of this objective. The numerical impact of this approximation will be examined in Section \ref{friedmanJ}.  

\begin{remark}
Though AIC was developed before TIC, it is easily reproduced as a special case of TIC. Subject to certain conditions (guaranteed by the requirements of  \cite{Akaike1973}), at least in expectation, $ I_{\mymletheta} = J_{\mymletheta} $. Thus, the quantity within the TIC trace term, $ I_{\mymletheta}J_{\mymletheta}^{-1} $, is the identity matrix. Therefore, its trace is  equal to $ p $, the parameter count of the model, recovering AIC. TIC itself can be derived using a proof that is similar to, though somewhat simpler than, the one we include in (\ref{biasTheorem}), of which Takeuchi's proof is a special case that is valid only at the MLE estimate $ \mymletheta $. 
\end{remark}

We also define $ \hat{J}^{*} $ to be the negative hessian of $ -\ell^{*}(\mytheta) $ rather than $ -\ell(\mytheta) $, and similarly for $ \hat{I}^{*} $, with expectations written as $ J^{*} $ and $ I^{*}$. Analogously, $ -\mathcal{L}^{*}(\mytheta) $ is the expectation of $ -\ell^{*}(\mytheta) $ and $ \myregtheta $ is the minimizer of $ -\ell^{*}(\mytheta) $, while $ \myregtheta_{0} $ is the minimizer of $ -\mathcal{L}^{*}(\mytheta) $.

We refer to the estimation of $\myregtheta$, by minimization of this corrected likelihood function as Information-Corrected Estimation (ICE). As the terminology suggests, we depart from the corrective approach used in Information Criterion, by directly minimizing the bias corrected likelihood function. Note that unlike $ L_{2} $ regularization, the correction term is parameter-free and thus would not require cross validation to estimate a hyperparameter such as the $ \lambda $ used by $ L_{2} $. 
 
General properties of this estimator are proved, and a set of regularity conditions are provided such that the estimator is asymptotically normal, and produces a bias that is $ O_{p}(n^{-3/2}) $ instead of the usual $ O_{p}(n^{-1}) $. Though this adds only a half-order to the bias correction, for most problems with reasonably large $ n $, any increase in order is likely to greatly reduce bias. Experimental results demonstrate superior properties of ICE for linear models compared to MLE with and without $L_2$ regularization.

\begin{remark}
For models satisfying White's regularity conditions (See \cite{White1982}), it is known that $ J_{\mytheta_0} $ is positive definite (thus non-singular) and continuous, and also that $ I_{\mytheta_0} $ is continuous with respect to $ \mytheta $. Therefore, $ \frac{1}{n}tr(I_{\mytheta_0}J_{\mytheta_0}^{-1}) $ would always be well defined in an open region around $ \mytheta_0 $. Similarly, the solution $ \myregtheta $ would be expected to have the same properties, and hence (for large enough $ n $) the estimate $ \frac{1}{n}tr(\hat{I}_{\myregtheta} \hat{J}_{\myregtheta}^{-1}) $ would be well defined when computed using the estimates $ \hat{I}_{\myregtheta} $ and $ \hat{J}_{\myregtheta}$. 

\end{remark}

\begin{remark}

N.B: Though $ -\ell^{*}(\mytheta) $ is an estimator of $ \mathcal{L}(\mytheta) $ accurate to within $ O(n^{-\frac{3}{2}}) $, that does not mean that $ \mathcal{L}(\myregtheta) $ is reduced by any particular amount relative to $ \mathcal{L}(\mymletheta) $. We expect that using this corrected objective will always (if it can be calculated accurately) generate some improvement by virtue of more accurately representing the true performance of the model out of sample, but there is no proof that this level of improvement has any particular form or asymptotic behavior. 

\end{remark}

Our approach preserves the linear complexity of training with respect to $n$. However, the computation of  $\hat{J}_{\myregtheta}^{-1}$ at each iteration of the numerical solver requires the inversion of a symmetric positive definite matrix with a complexity of $O(p^3)$. Hence the approach is not suitable for high dimensional datasets without adjustment. See Section \ref{sect:optObj} for optimized approximations that are viable for larger parameter counts. Further exploration of large models based on this approach are beyond the scope of the present work.

\begin{remark}
It is clear from inspection that if $ -\ell(\mytheta) $ is strictly convex, then so too is $ -\ell^{*}(\mytheta) $ for large enough $ n $. 
\end{remark}

We first provide a proof of asymptotic convergence of $\myregtheta$ under certain regularity conditions. With this convergence result in place, we then show that minimizing \eqref{eq:ice} leads to an $O(n^{-3/2})$ bias term, an improvement over the $ O(\frac{1}{n}) $ term produced by MLE.


\subsection*{Local Behavior of the ICE Objective}

Suppose the following conditions hold:

\begin{enumerate}
\item $\mathcal{M}$ satisfies White's regularity conditions A1--A6 (see Section \ref{sect:white} or \cite{White1982}). 
\item $ \mytheta_0 $ is a global minimum of $ -\mathcal{L}(\mytheta) $ in the compact space $ \Theta $ defined in A2. 
\item There exists a $ \varepsilon > 0 $ such that $ -\mathcal{L}(\mytheta_0) < -\mathcal{L}(\mytheta_1) - \varepsilon $ for all other local minima $ \mytheta_1 $.
\item For $ k = {0, 1, 2, 3, 4, 5} $ the derivative $ \partial^{k}_{\mytheta} \mathcal{L}(\mytheta) $ exists, is continuous, and bounded on an open set around $ \mytheta_0 $. 
\item For $ k = {0, 1, 2, 3, 4, 5} $, the variance $ \myV[\partial^{k}_{\mytheta} \ell(\mytheta, \Var_n)] \rightarrow 0 $ as $ n \rightarrow \infty $ on an open set around $ \mytheta_0 $. 
\end{enumerate}

Then for sufficiently large $ n $ there exists a compact subset $ U \subset \Theta $ containing $ \mytheta_0, \mymletheta, $ such that:

\begin{enumerate}
\item For $ k = {0, 1, 2, 3} $ the derivative $ \partial^{k}_{\mytheta} \ell^{*}(\mytheta, \data) $ exists, is continuous, and bounded on $ U $, almost surely. 
\item For $ k = {0, 1, 2, 3} $, $ \myV[\partial^{k}_{\mytheta} \ell^{*}(\mytheta, \Var_n)] \rightarrow 0 $ as $ n \rightarrow \infty $ on $ U $, almost surely. 
\item $ \myregtheta \in U $ as $ n \rightarrow \infty $ almost surely. 
\item $ \sqrt{n}(\myregtheta -\myregtheta_0)  \rightarrow N(0,  (\hat{J}_{\myregtheta_{0}}^{*})^{-1}\hat{I}^{*}_{\myregtheta_{0}}(\hat{J}_{\myregtheta_{0}}^{*})^{-1}) $ almost surely.
\item $ -\mathcal{L}(\hat{\myregtheta}) =  -\ell^{*}(\myregtheta(\Var_{n}), \Var_{n}) + O_{p}(n^{-3/2}) $ almost surely.
\end{enumerate}

Items (1--3) follow from Lemma \ref{asymptoticLemma} (see Section \ref{sect:fvproof}). These are additional regularity conditions that are prerequisites for later theorems. 

Item (4) follows from Theorem \ref{normalityTheorem} in Section \ref{sect:anproof}. This states that the estimate $ \myregtheta $ is asymtotically normal in a way that is analogous to classical asymptotic normality results for MLE. It is only true almost surely because results (1--3) upon which it relies are only true almost surely. 

Item (5) follows from Theorem \ref{biasTheorem} in Section \ref{sect:pboproof}. This item establishes the superior accuracy of the ICE objective compared to the MLE objective function in predicting out of sample errors. Like item (4) this is only true almost surely because intermediate results on which it relies are only true almost surely.

The reduction in generalization error seen arises from the optimization over the superior ICE objective function, analogous to the way that $ L_{2} $ regularization is used for this purpose. 

\begin{remark}
The regularity conditions described here are only slightly more strict than the conditions described by White \cite{White1982}. In particular, models having three continuous derivatives as required by White, but not 5 as needed here are thought to be very rare. Requirement (2) is just the definition of $ \mytheta_0 $, which White labels differently, and requirement (3) excludes a pathological corner case, the further study of which is beyond the scope of this paper. 
\end{remark}

\begin{remark}
Note that as $ -\ell(\mytheta, \data) $ is convex in the neighborhood of $ \mytheta_0 $, so too is $ -\ell^{*}(\mytheta) $ for large enough $ n $ because $ -\ell^{*}(\mytheta) \rightarrow -\ell(\mytheta) $.
Thus it can be concluded that the local behavior of $ -\ell^{*} $ in the neighborhood of $ \mytheta_0 $ is not appreciably worse than the behavior of $ -\ell $ if the problem is not too ill conditioned. 
\end{remark}


\section{Direct Computation Results}
\label{sect:direct}

The following experiments have been designed to compare MLE, MLE with $ L_{2} $ regularization, and ICE for regression. Each experiment involves simulation of training and test sets and is implemented in R. See the attached code to run each experiment. 

Each of these experiments has been performed using the raw formula for ICE provided in Equation (\ref{eq:ice}) with minimal adjustments. All gradients are computed using R's default finite difference approach. This means that for a model with $ p $ parameters, the objective function is dominated by the inversion of $ J $, which costs $ O(p^{3}) $ time and $ O(p^{2}) $ space. The use of finite difference gradients further increases the time complexity to $ O(p^{4}) $, compounding the problem. This approach is therefore viable for small models with few parameters, but not realistic for larger models. Optimizations to overcome this limitation will be considered in upcoming Section \ref{sect:optObj}. The use of finite difference derivatives was not found to produce appreciable numerical differences in the final output, so analytic derivatives were not used for this analysis. 

The code and results for this section is provided in \cite{ward_2021}. Throughout this section, the following estimators will be compared.

\begin{table}[H]
\begin{center}
\begin{tabular}{cccc}
\hline
MLE &  $ \mymletheta(\newVar_n) $ & $ := \argmin_{\mytheta} [-\ell(\mytheta)] $ \\
$L_2$ regularization & $\myregtheta_{L_{2}}(\newVar_n) $ & $ := \argmin_{\mytheta, \lambda} [-\ell(\mytheta) + \lambda \| \mytheta \|^{2}_{2}] $ \\
ICE &  $\myregtheta_{ICE}(\newVar_n) $ &:= $ \argmin_{\mytheta} [-\ell(\mytheta) +  \frac{1}{n}tr(\hat I_{\mytheta} \hat J^{-1}_{\mytheta})]$ \\
\hline
\end{tabular}
\end{center}
\end{table}


\subsection{Gaussian Error Model}
\label{gaussian_example}

We begin by considering the simplest case of univariate linear regression with Gaussian residuals.  The advantage of this simple model is that the exact form of the correction term can be derived analytically and aids therefore in building intuition on its behavior. For such a toy model, $ y \sim N(\mu, \sigma^{2}) $ and, for simplicity, the following example will consider $ \mu $ to be a constant, but it is equally applicable if $ \mu = \mu(\myx) $. Consider the parameters of the model to therefore be $ \mytheta := (\mu, \sigma) $ with their optimal values being $ \mytheta_0 := (\mu_{0}, \sigma_0) $. The the probability density function is 

\begin{equation}
g(y, \mytheta) = \frac{1}{\sqrt{2 \pi \sigma}} e^{-\frac{(y - \mu)^{2}}{2\sigma^{2}}}.
\end{equation}

It is known a priori that $ L_{2} $ regularization cannot improve this model, as if $ \mu_{0} \neq 0 $, any decrease in the magnitude of $ \mu $ is likely to be systematically harmful. Similarly, a decrease in $ \sigma $ below $ \hat{\sigma} $ results in a decrease in model distribution entropy, and hence would be generally making overfitting worse, and would generate a correspondingly higher KL-divergence than the MLE estimate. Consequently, we would expect any $ \lambda $ computed through cross-validation to be statistically indistinguishable from zero, and $ L_{2} $ regularization to be generally harmful whenever $ \lambda \neq 0 $. 

\subsubsection*{Generalization Error Analysis}

The Gaussian model described was generated with $ \mu_{0} = 0.2 $, $ \sigma_{0} = 0.2 $, and $ dy = 0.001 $. For each of $ n \in \{16, 32, 64, 128, 256, 512, 1024 \} $, $ 500 $ independent simulations of the data $ y_{1}, \dots, y_{n} $ were performed, and then the parameters were fit from that data. In each simulation, $ \mytheta $ was computed using MLE, MLE with $ L_{2} $ regularization, and ICE. The $ \lambda $ parameter for $ L_{2} $ regularization was computed using 2-way cross-validation on the available data, and as expected, none of the computed values of $ \lambda $ were statistically different from zero. 

For each estimate of $ \mytheta $, the KL-divergence $ \rho_{KL}(f, g_{\mytheta}) $ was computed (using the known value of $ \mytheta_0 $), and the results were compared. The ICE parameter estimation method showed statistically significant improvement over MLE at the 5-sigma level out to $ n = 64 $, and was improved by just under 1-sigma at $ n = 1024 $. 

\begin{figure}[H]
\includegraphics[width=0.65\linewidth]{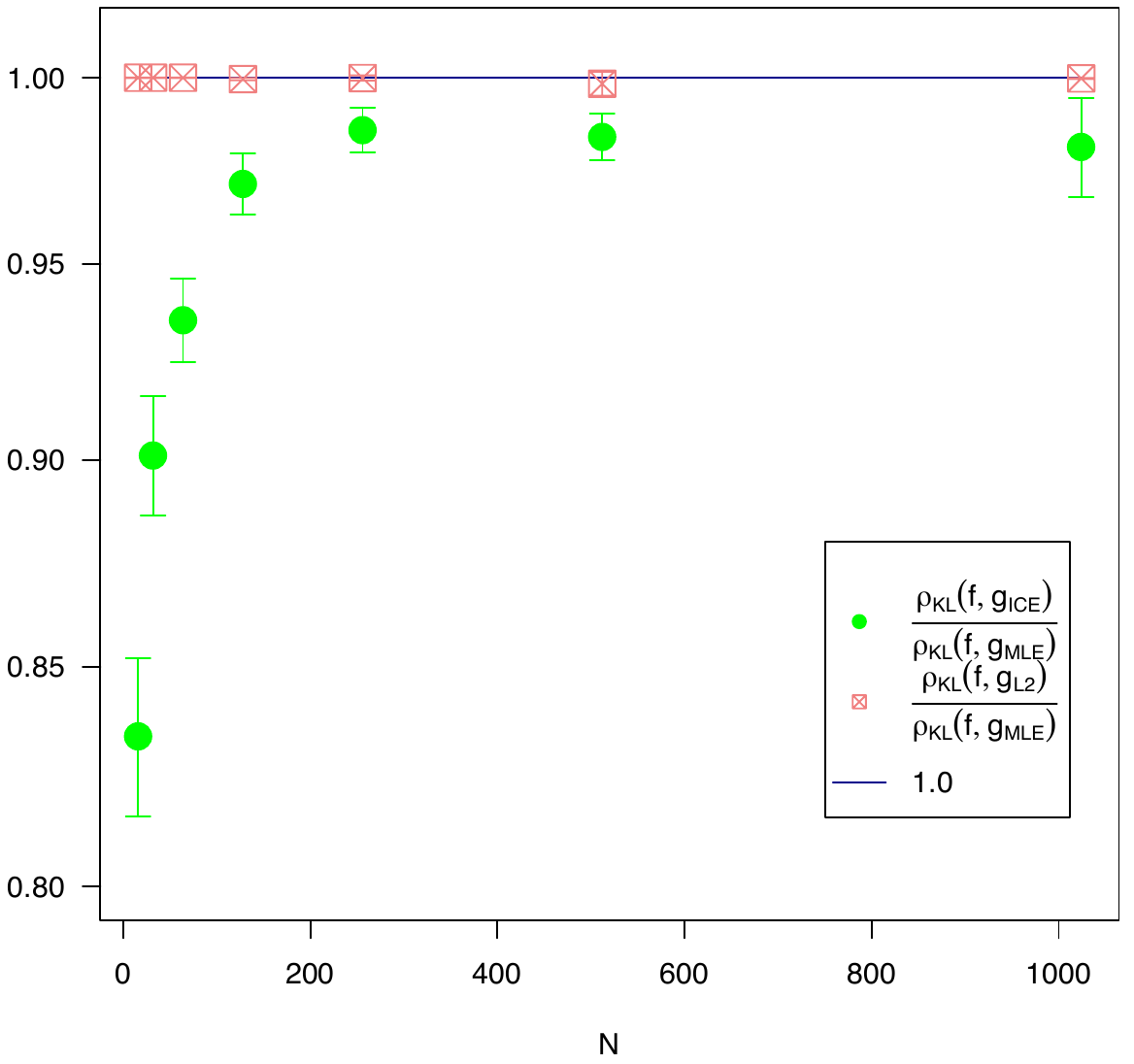}
\caption{{A comparison of the KL-divergence (y-axis) of various estimation methods against the number of training samples $n$. Each KL divergence value was divided by the average KL divergence of the MLE estimate for that value of $ n $. The ICE and $ L_{2} $ series are shown with 2 standard deviation error bars.}}
\label{gaussianKL}
\end{figure}

The KL-divergence results graphed against $ n $ on a log-log scale are shown in Figure \ref{gaussianKL}. Every value of $ n $ is normalized by the average KL divergence of the MLE methodology to improve legibility. The $ L_{2} $ series is statistically indifferent from the MLE series at $ 2 $ standard deviations beyond $ n = 32 $, and the two are not materially different for any $ n $. The ICE series is at least $ 4.5 $ standard deviations below the MLE series until $ n = 1024 $.

\begin{remark}
In addition to the series shown in Figure \ref{gaussianKL}, a series was computed using the true value of $ J $, estimated from a much larger sample $ n=1024 $ from the underlying distribution, and this series was indistinguishable from the series computed using $ \hat{J} $ for every $ n $, thus it was not graphed. This validates Takeuchi's approach of approximating $ J $ with $ \hat{J} $ in this instance.
\end{remark}

\begin{figure}[H]
\includegraphics[width=0.65\linewidth]{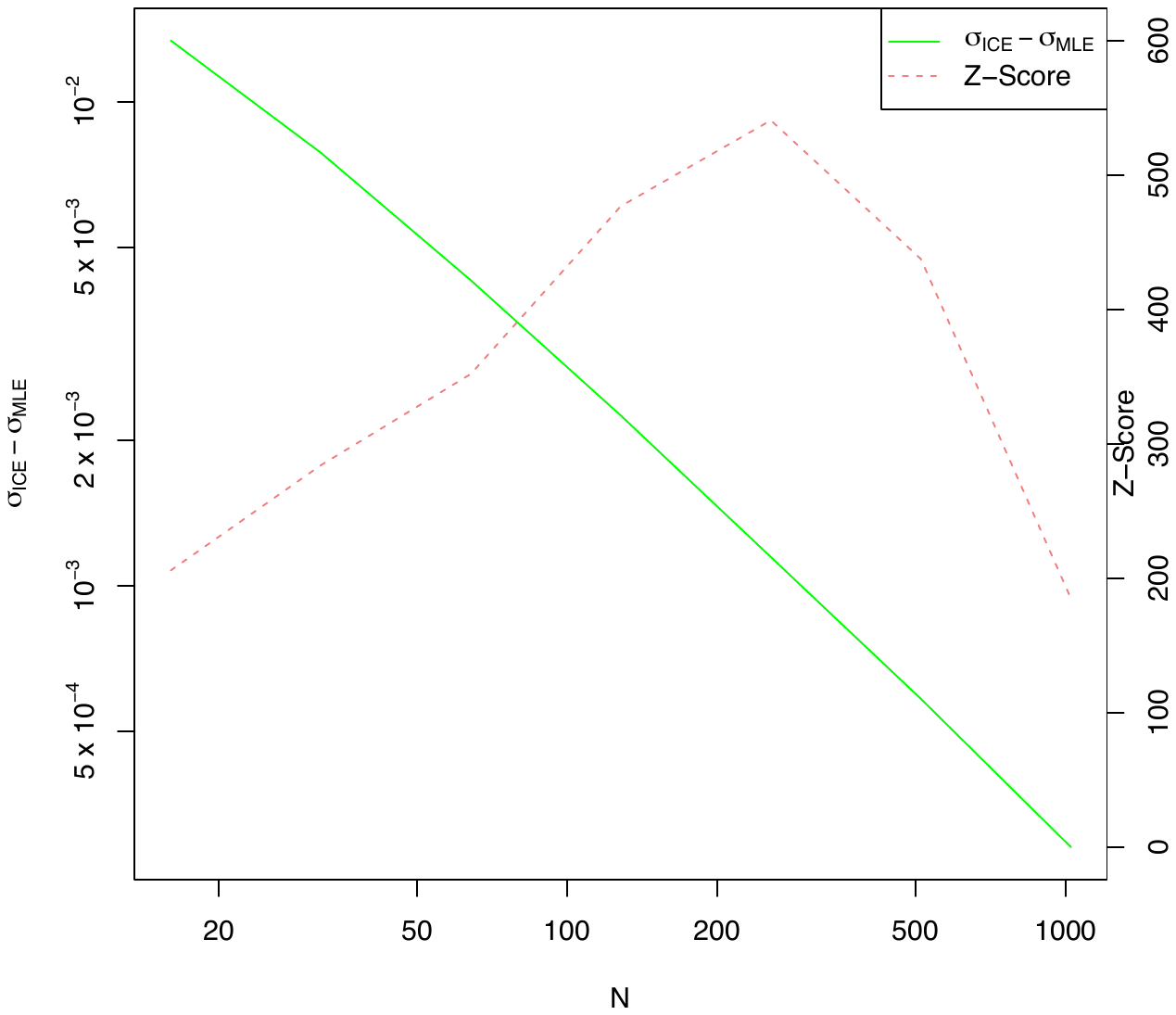}
\caption{{The error in the estimated $\hat{\sigma}_{ICE}$ and the Z-score of the estimate against the number of training samples $n$.}}
\label{gaussianSigma}
\end{figure}

As expected, the difference in $ \mu $ between ICE and MLE is not statistically significant (at three standard deviations) for any $ n $, but the ICE computed value of $ \sigma $ (shown in Figure \ref{gaussianSigma}) is considerably larger than the MLE estimate, especially for small values of $ n $. This explains the greatly reduced KL-divergence noted in Figure \ref{gaussianKL}. 

Note that the difference in estimated $ \sigma $ is always statistically significant when compared to the MLE value. This is because both MLE and ICE are fit on the same data, so ICE would always have a larger $\sigma$ than MLE regardless of the actual data chosen from the distribution $ f $. This is the cause of the large z-scores shown, always exceeding 200. We know from elementary statistics that correlation between the mean and std. deviation causes the MLE estimate of $ \hat{\sigma} $ to be systematically low by a factor of $ \frac{n-1}{n} $. Indeed, the ICE estimate of $ \sigma^{*} $ is closely tracking $ \sigma_0 $ whereas $ \hat{\sigma} $ is closely tracking $ \frac{\sigma_0 (n-1)}{n} $ as expected. This is one example where reducing generalization error also reduces parameter bias as a side effect. 


\subsection{Friedman's Test Case}
\label{FriedmanTest} 

We now extend the example from Section \ref{gaussian_example} to the case where $ \mu $ is no longer constants. For this example, we chose a standard regression test set, which is nonlinear in the features, based on Section 4.3 of \cite{MARS}:

\begin{equation}
y_{i} = \mu_{\mytheta}(x_{i}) + \varepsilon_{i}, ~\varepsilon\sim N(0,\sigma^2),
\end{equation}
where the Friedman model is

\begin{equation}
\mu_{\mytheta}(x_{i}) = \theta_0 sin(\pi x_{(i, 0)}x_{(i, 1)}) + \theta_1 (x_{(i, 2)} - \theta_2)^2 + \theta_3 x_{(i, 3)} + \theta_4 x_{(i, 4)}.
\end{equation}

The random features, $ X_{j} $, are i.i.d. uniform random and the parameter values are fixed.  The true parameter set,  $\mytheta_0 = (10.0, 20.0, 0.5, 10.0, 5.0, 1.0)$, reserves the last parameter ($ 1.0$) for the value of $ \sigma $. 

Note that here $ \sigma $ must be treated as an unknown parameter. To do otherwise implies that the modeler knows the amount of noise expected in the data. In the case of a known noise term, overfitting is impossible since overfitting arises when a model reduces the projected noise below its actual value, which can never arise when the noise level is known. 

The model probability density $ g(x, y |  \mytheta) $ of $ y $ is given by

\begin{equation}
g(x_{i}, y_{i} | \mytheta) = \frac{1}{\sqrt{2\pi \sigma^{2}}} e^{-\frac{(\mu_i- y_{i})^{2}}{2 \sigma^{2}}}.
\end{equation}

Recall that in Section \ref{gaussian_example}, the value of $ \mu $ was considered to be a constant. This example is a natural extension of Section \ref{gaussian_example}, and was chosen due to the well-explored difficulty of Friedman's problem. 

We simulate $ 500 $ batches of equally sized training sets of length 
$ n \in \{16, 32, 64, 128, 256,$ $ 512, 1024 \}$. The test set is always of length $ 1024 $ to ensure accuracy for the smaller values of $ n $. The starting point of the optimization is generated by adding a random perturbation, $ \delta \mytheta \sim N(0,0.1) $, to each parameter. As before, the KL-divergence is computed between the distribution represented by the parameters and the true distribution, and these values are compared between estimation methods. 

For each test sample, the KL divergence is computed using numerical integration with a $ dy $ increment of $ 0.01 $ over the interval containing $ \mu \pm 10 \sigma $ for both the true and model distributions. The computed probabilities are verified to numerically sum to unity within an error of $\pm 10^{-3} $. 

In each simulation, $ \mytheta $ is computed using MLE, MLE with $ L_{2} $ regularization, and ICE. The $ \lambda $ parameter for $ L_{2} $ regularization is computed using 4-way cross validation on each batch of the training data.



\begin{figure}[H]
 \includegraphics[width=0.65\linewidth]{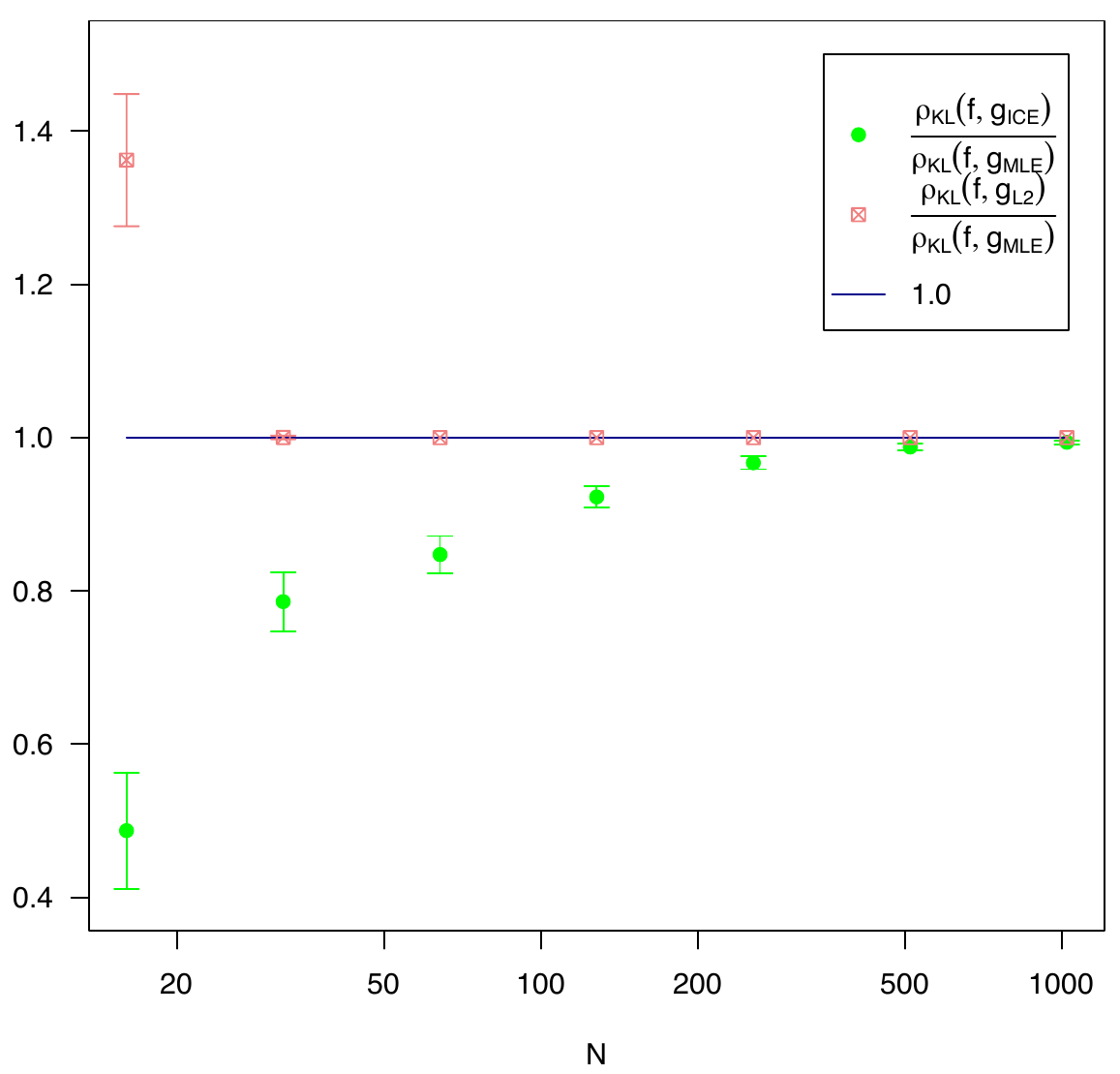}
\caption{{Comparison of the KL-divergence, averaged across $ 500 $ replications, of estimation methods against the number of training samples $n$. Each KL divergence value was divided by the average KL divergence of the MLE estimate for that value of $ n $. The ICE and $ L_{2} $ series are shown with 2 standard deviation error bars.}}
\label{friedmanKL}
\end{figure}

As shown in Figure \ref{friedmanKL} and Table \ref{tab:friedman}, $ L_{2} $ is not effective for any value of $ n $, and is is completely inactivated for $ n > 32 $. Where regularization is used (i.e., $ \lambda \neq 0 $), it generally underperforms MLE. ICE is effective across the entire data range, outperforming MLE for every $ n $, and always by a statistically significant margin of at least 5 sigma. 

\begin{table}[!h]

\begin{center}
\begin{tabular}{cccc}
\hline
\textbf{\emph{n}} & \boldmath{$ \rho_{KL}(f, g_{\mymletheta}) $}  & \boldmath{$ \rho_{KL}(f, g_{\mytheta_{L2}}) $}  & \boldmath{$ \rho_{KL}(f, g_{\myregtheta}) $}    \\
\hline
$ 16 $ & $ 6.19 \times 10^{-1} $ & $ 8.442 \times 10^{-1} $ (8.40) & $ 3.02 \times 10^{-1} $ ($-$13.54) \\
$ 32 $ & $ 1.74 \times 10^{-1} $ & $ 1.74 \times 10^{-1} $ (0.16) & $ 1.37 \times 10^{-1} $ ($-$11.11) \\
$ 64 $ & $ 6.85 \times 10^{-2} $ & $ 6.85  \times 10^{-2} $ (0.0) & $ 5.81 \times 10^{-2} $ ($-$12.65) \\
$ 128 $ & $ 3.82 \times 10^{-2} $ & $ 3.82  \times 10^{-2}$ (0.0) & $ 3.53 \times 10^{-2} $ ($-$11.13) \\
$ 256 $ & $ 2.19 \times 10^{-2} $ & $ 2.19  \times 10^{-2}$ (0.0) & $ 2.12 \times 10^{-2} $ ($-$7.84) \\
$ 512 $ & $ 1.53 \times 10^{-2} $ & $ 1.53  \times 10^{-2}$ (0.0) & $ 1.51 \times 10^{-2} $ ($-$5.66) \\
$ 1024 $ & $ 1.25 \times 10^{-2} $ & $ 1.25  \times 10^{-2}$ (0.0) & $ 1.24 \times 10^{-2} $ ($-$5.03) \\
\hline
\end{tabular}
\caption{{Comparison of the average KL divergence across $ 500 $ replications for several model estimators given a fitting set size of $ n $. For estimators other than $ \mymletheta $, the values in parentheses denotes the t-statistic of the difference between this estimator and $ \mymletheta $, with negative values indicating that the listed estimator has a lower KL divergence.}}
\label{tab:friedman}
\end{center}
\end{table}

\subsubsection{Impact of $ \hat{J} $ Approximation}
\label{friedmanJ} 

It was noted previously that Takeuchi used $ \hat{J} $ (and likewise, $ \hat{I}$) in place of the true value of $ J $, and we do so here as well. Though there is no realistic way to avoid this approximation in the real world, and the optimized approach discussed in Section \ref{sect:optObj} has an entirely different set of approximations, the impact of this approximation will be briefly characterized here. 

In Table \ref{tab:friedmanTrue}, we revisit Table \ref{tab:friedman}, but now drop the $ L_{2} $ regualarization column, and add a new column where the ICE objective is allowed to use a much better approximated value of $ J $, in this case approximated from $ 1024 $ independently drawn samples regardless of $ n $. 

\begin{table}[!h]

\begin{center}
\begin{tabular}{cccc}
\hline
\textbf{\emph{n}} & \textbf{MLE} & \textbf{ICE (\boldmath{$\hat{J}$} )}  & \textbf{ICE (\boldmath{$J$} ) }   \\
\hline
$ 16 $ & $ 6.19 \times 10^{-1} $ & $ 3.02 \times 10^{-1} $ ($-$13.54) & $ 6.21 \times 10^{-1} $ (0.05) \\
$ 32 $ & $ 1.74 \times 10^{-1} $ & $ 1.37 \times 10^{-1} $ ($-$11.11) & $ 1.51 \times 10^{-1} $ ($-$3.74) \\
$ 64 $ & $ 6.85 \times 10^{-2} $ & $ 5.81  \times 10^{-2} $ ($-$12.65) & $ 4.21 \times 10^{-2} $ ($-$17.29) \\
$ 128 $ & $ 3.82 \times 10^{-2} $ & $ 3.53  \times 10^{-2}$ ($-$11.13) & $ 2.99 \times 10^{-2} $ ($-$11.82) \\
$ 256 $ & $ 2.19 \times 10^{-2} $ & $ 2.12  \times 10^{-2}$ ($-$7.84) & $ 1.99 \times 10^{-2} $ ($-$4.19) \\
$ 512 $ & $ 1.53 \times 10^{-2} $ & $ 1.51  \times 10^{-2}$ ($-$5.66) & $ 1.48 \times 10^{-2} $ ($-$1.62) \\
$ 1024 $ & $ 1.25 \times 10^{-2} $ & $ 1.24  \times 10^{-2}$ ($-$5.03) & $ 1.24 \times 10^{-2} $ ($-$0.72) \\
\hline
\end{tabular}
\caption{{Comparison of the average KL divergence across $ 500 $ replications for ICE estimators with and without approximation of $ J $ given a fitting set size of $ n $. For estimators other than $ \mymletheta $, the values in parentheses denotes the t-statistic of the difference between this estimator and $ \mymletheta $, with negative values indicating that the listed estimator has a lower KL divergence.}}
\label{tab:friedmanTrue}
\end{center}
\end{table}

As can be seen from Table \ref{tab:friedmanTrue}, using the true value of $ J $ is at most marginally helpful. In fact, for most values of $ n $ it displays slightly better average results, but slightly higher std. deviation of those results, and thus reduced T-statistics. Thus, we conclude that the Takeuchi's approximation, replacing $ J $ with $ \hat{J} $ is reasonable. The same conclusion was reached in Section \ref{gaussian_example}, see the remark there. We note also that the ICE estimator using $ \hat{J} $ exhibits substantially better performance for very low sample sizes, but further investigation of this phenomenon is beyond the scope of the current paper.

In Table \ref{normTab}, we show the average matrix norms of $ J $, $ \hat{J} $, and also of the diagonal of $ \hat{J} $, referred to as the matrix $ D $. The matrix $ D $ will be examined further in Section \ref{sect:optObj}, and is included here for completeness. We also show the norms of several matrix differences.

We note that the ICE objective values themselves exhibit much lower variation than the matrix norms show in Table \ref{normTab}. In particular though the matrix $ D $ is not actually converging to $ J $ as $ n $ increases, we see from the correction term it generates that this difference does not appear to have a material impact for larger $ n $. We thus conclude that the major eigenvectors of $ (\hat{J} - J)  $ and $ (D - J) $ are very nearly orthogonal to the gradient vectors used to construct $ \hat{I} $ for large $ n $.

\clearpage

\begin{table}[!h]
\begin{center}
\begin{tiny}
\begin{tabular}{ccccccccc}
\hline
\textbf{\emph{n}} & \boldmath{$ \|J \| $}  &  \boldmath{$ \|\hat{J} \| $}  & \boldmath{$ \|D  \| $}  & \boldmath{$ \| \hat{J} - J  \| $}   & \boldmath{$ \| D - J \| $}  & \boldmath{$ \frac{1}{n} tr(IJ^{-1}) $}   & \boldmath{$ \frac{1}{n} tr(\hat{I}\hat{J}^{-1}) $}  & \boldmath{$ \frac{1}{n} tr(\hat{I}D^{-1}) $}  \\
\hline
 16  & 1423.29  &  925,489.26  &  1738.30  &  926,165.93 &  2905.64  &  0.1452  &  0.0027  &  6.9484  \\
 32  &  875.03  &  89,414.11  &  96.49   &  89,786.19 &  793.43   &  0.0521  &  0.0256  &  0.0291  \\
 64  &  214.05  &  4820.55  &  89.55   &  4646.51 &  125.02   &  0.0202  &  0.0160  &  0.0162  \\
 128  &  200.86  &  200.68  &  85.00  &  27.65 &  115.86   &  0.0097  &  0.0086  &  0.0086   \\
 256  &  194.36  &  191.81  &  82.70   &  19.12 &  111.67   &  0.0048  &  0.0045  &  0.0045  \\
 512  &  191.20  & 190.10  &  82.76   &  14.18 &  108.44   &  0.0023  &  0.0023  &  0.0023  \\
 1024  &  188.91  &  187.39  &  81.89  &  11.58 &  107.02   &  0.0012  &  0.0012  &  0.0012   \\
\hline
\end{tabular}
\caption{{Mean matrix norms of $ J $, its approximations, and differences from these approximations across 500 replications.}}
\label{normTab}
\end{tiny}
\end{center}
\end{table}

It is not clear from examining the trace terms in Table \ref{normTab} that $ D $ is a worse approximation of $ J $ than $ \hat{J} $ is, even for small $ n $ where the impact of the ICE approach is most significant.  A more complete investigation of the spectrum of these matrices is beyond the scope of the present work.


\subsection{Multivariate Logistic Regression} 
\label{sect:results}

The previous experiment is based on a well-known test case. In this second experiment, we assess the general performance of ICE under (i) varying dimensionality of the true data distribution, (ii) increasing misspecification, and (iii) increasing training set sizes. To achieve this goal, we generate a more exhaustive set of data from a more complex data generation process.

\subsubsection{Data Generation Process}
The synthetic data are designed to exhibit a number of characteristics needed to broadly evaluate the efficacy of ICE. First, the regressors should be sufficiently correlated so as to ensure that model selection is representative of typical datasets. However, we avoid multi-collinearity by ensuring the smallest eigenvalue is above a certain threshold. We additionally control the condition number of the covariance matrix $ \Sigma $ by randomly generating a symmetric positive definite covariance matrix $\Sigma\in\mathbb{R}^{p}$ using the eigen-decomposition
\be
\Sigma=UDU^T,
\ee
where $U$ is an orthogonal random matrix with elements $U_{ij}\sim N(0,1)$ and $D$ is diagonal matrix of positive eigenvalues. The eigenvalues are uniformly distributed over the interval $[a,b]$ so that the condition number of $\Sigma$ is $b/a$ and the eigenvalues are kept distinct. Here, $a$ is chosen to be $1\times 10^{-4}$ and $b$ is chosen to be 0.1.

Using a Cholesky decomposition  $\Sigma= \Gamma\Gamma^T$ and the random mean vector $\mu \sim N(0,1)$, we generate correlated gaussian vectors of dimension $ p $ with the properties
\begin{equation}
X_{i} = \mu + \Gamma_{ij} Z_{j},~ Z_j\sim N(0,1), \forall j \in{1,\dots, p}.
\end{equation}

The data $(\data, y_n)$ are generated under a logistic regression
\begin{equation}
p(y =  1 | \myx, \mytheta_0) = f(\myx | \mytheta_0) = \frac{1}{1+ e^{-\myx \mytheta_0}}.
\end{equation}
A key challenge in assessing the efficacy of bias reduction is to avoid generating excessively low entropy distributions. In such cases, bias reduction will have marginal effect as the parameters are all nearly zero. To avoid such scenarios, the intercept parameter of the true model is adjusted a-posterior until the following conditions are met:
\begin{enumerate}
\item $ c < \myE_{\newVar}[p(Y = 1 | \myX, \mytheta_{0})] < d $
\item $ -\mathcal{L}(\mytheta_{0}) > \epsilon $
\end{enumerate}
where $c=0.35$, $d=0.65$, and $\epsilon=0.2$. If these conditions can not be met, then the replication is discarded.

\subsubsection{Model Performance Comparison}

As in prior sections, KL divergence is computed between the estimated model and the true model for each of the estimation methods. The T-statistics of the difference with the corresponding MLE KL divergence are computed, with negative T-statistics showing that an approach is performing better than the MLE approach. For $ L_{2} $ regularization in this section, the value of $ \lambda $ is computed via cross-validation, using two folds, on the provided fitting set. 

Table  \ref{tab:nonlint} compares the KL divergences $ \rho_{KL} $ from the true distribution to the model distributions produced using various estimation approaches applied to misspecified data. Here $ m $ denotes the number of regressors that are not predictive, i.e., $ \mytheta_0 $ contains $m$ zeros.  The experiment is replicated $300$ times using the data generation process described above and the test set is fixed at 100,000 observations.

\begin{table}[!h]
\begin{tabular}{cccccc}
\hline
\textbf{\emph{p}} &   \textbf{\emph{n}} & \boldmath{$ \rho_{KL}(f, g_{\mymletheta}) $}  & \boldmath{$ \rho_{KL}(f, g_{\mytheta_{L_{2}}}) $}   &  \boldmath{$ \rho_{KL}(f, g_{\myregtheta}) $}   \\
\midrule
5 &   500 & $ 4.79 \times 10^{-3} $  & $ 3.01 \times 10^{-3} $ ($-$13.43)  & $ 4.56 \times 10^{-3} $ ($-$13.53) \\
5 &   1000 & $ 2.64 \times 10^{-3} $  & $ 1.76 \times 10^{-3} $ ($-$10.94)  & $ 2.57 \times 10^{-3} $ ($-$12.80) \\ 
5 &   2000 & $ 1.29 \times 10^{-3} $  & $ 1.09 \times 10^{-3} $ ($-$6.15)  & $ 1.27 \times 10^{-3} $ ($-$7.69)  \\
5 &  5000 & $ 5.09 \times 10^{-4} $  & $ 4.60 \times 10^{-4} $ ($-$4.69)  & $ 5.07 \times 10^{-4} $ ($-$6.19)  \\
\midrule
10 &  500 & $ 9.79 \times 10^{-3} $ & $ 9.85 \times 10^{-3} $ (0.16)  & $ 9.18 \times 10^{-3} $ ($-$6.27)  \\
10 &  1000 & $ 5.05 \times 10^{-3} $ & $ 5.13 \times 10^{-3} $ (0.51)  & $ 4.83 \times 10^{-3} $ ($-$4.90)  \\
10 &  2000 & $ 2.50 \times 10^{-3} $ & $ 3.05 \times 10^{-3} $ (5.99)  & $ 2.56 \times 10^{-3} $ (1.70)  \\
10 &  5000 & $ 1.06 \times 10^{-3} $ & $ 1.49 \times 10^{-3} $ (7.72)  & $ 1.04 \times 10^{-3} $ ($-$0.86)  \\
\midrule
20 &  500 & $ 2.18 \times 10^{-2} $ & $ 2.16 \times 10^{-2}  $($-$0.29)  & $ 1.95 \times 10^{-2} $ ($-$8.71)  \\
20 &  1000 & $ 1.13 \times 10^{-2} $ & $ 1.24 \times 10^{-2} $ (3.79)  & $ 1.10 \times 10^{-2} $ ($-$1.95)  \\
20 &  2000 & $ 6.86 \times 10^{-3} $ & $ 7.52 \times 10^{-3} $ (4.47) & $ 6.72 \times 10^{-3} $ ($-$1.67)  \\
20 &  5000 & $ 3.57 \times 10^{-3} $ & $ 4.24 \times 10^{-3} $ (6.56)  & $ 3.59 \times 10^{-3} $ (0.45)  \\
\hline
\end{tabular}

\caption{{Comparison of the KL divergence for the different estimation approaches applied to mis-specified data. The values in parentheses denote the t-statistic relative to MLE. For $ p = \{5, 10, 20\} $ there are $ m = \{2, 4, 8 \} $ non-explanatory variables added. }}
\label{tab:nonlint}
\end{table}


We observe that the t-statistic for $ \myregtheta $ is most significant for relatively small sample sizes, particularly $ n = 500 $. For these small sizes, the improvement over MLE is greater, though noisier. There is uniform decay in improvement over $ \mymletheta $ as $ n $ grows, until for $ p = 10 $ and $ p = 20 $ the largest sizes are no longer statistically significant.  This is expected, as both the MLE and ICE estimates are converging towards the true value of $\mytheta_0$, and for large enough sample sizes the ICE correction would be dominated by numerical error, particularly the ill conditioning of $ J $.  

The $ L_{2} $ estimate improves for small values of $ p $, but then becomes progressively worse for large values of $ p $. We observe that for dimensionality above $ p=5 $, the $ L_{2} $ regularization described here is no longer effective in reducing the KL-divergence. For low values of $ p $ the value of $ \mytheta \myx $ has comparatively low variance, and thus the logistic function is reasonably locally approximated as linear. For higher $ p $ this approximation is less realistic and the performance of $ L_{2} $ regularization degrades.

For the ICE estimates, larger values of $ p $ show fluctuations that are often not statistically significant. It is apparent that larger $ p $ is increasing the variance of the ICE divergences, probably due to numerical errors and ill conditioning. Larger values of $ n $ reduce the absolute size of the divergence improvement whereas larger values of $ p $ seem to increase~it. 

Note that though the t-statistics are degrading for large $ n $, the absolute magnitude of the differences is asymptotically small. For these sizes, the results are insignificant, but more importantly, immaterial.  

\subsubsection{Convergence Analysis for Large \emph{n}}

For $ 10 $ randomly chosen example problems, under which the model coefficients are now fixed, the convergence behavior for large $ n $, the training set size, is explored. Note that the test set remains fixed at 100,000 observations for each problem. Table \ref{tab:converge} compares the KL divergence (averaged over all $ 10 $ problems) under MLE ($\mymletheta$), $L_2$ regularization, and ICE for progressively larger sample sizes. The divergences $  \rho_{KL}(f, g_{\mymletheta}) $ and $ \rho_{KL}(f, g_{\myregtheta_{ICE}})  $ converge to zero as $ n \rightarrow \infty $, as does $  \rho_{KL}(f, g_{\myregtheta_{L_{2}}})  $.

\begin{table}[H]
\begin{tabular}{cccccc}
\hline
 \textbf{\emph{n}} & \boldmath{$ \mathcal{L}(\mytheta_0) $}   & \boldmath{$ \rho_{KL}(f, g_{\mymletheta})  $}  & \boldmath{$ \rho_{KL}(f, g_{\mytheta_{L_{2}}}) $}  & \boldmath{$ \rho_{KL}(f, g_{\myregtheta}) $}  \\
\midrule
500 & $ 0.5439 $ & $ 9.28 \times 10^{-3} $ & $ 7.92 \times 10^{-3} $ & $ 7.74 \times 10^{-3} $  \\
1000 & $ 0.5439 $ & $ 5.50 \times 10^{-3} $ & $ 5.86 \times 10^{-3} $ & $ 4.81 \times 10^{-3} $ \\
2000 & $ 0.5439 $ & $ 2.65 \times 10^{-3} $ & $ 3.67 \times 10^{-3} $ & $ 2.65 \times 10^{-3} $ \\
5000 & $ 0.5439 $ & $ 1.85 \times 10^{-3} $ & $ 2.72 \times 10^{-3} $ & $ 1.35 \times 10^{-3} $  \\
10,000 & $ 0.5439 $ & $ 5.75 \times 10^{-4} $ & $ 1.57 \times 10^{-3} $ & $ 9.32 \times 10^{-4} $ \\
20,000 & $ 0.5439 $ & $ 5.84 \times 10^{-4} $ & $ 8.10 \times 10^{-4} $ & $ 6.11 \times 10^{-4} $  \\
50,000 & $ 0.5439 $ & $ 3.83 \times 10^{-4} $ & $ 4.09 \times 10^{-4} $ & $ 3.64 \times 10^{-4} $ \\
100,000 & $ 0.5439 $ & $ 1.67 \times 10^{-4} $ & $ 1.15 \times 10^{-3} $ & $ 1.86 \times 10^{-4} $ \\
\hline
\end{tabular}
\caption{{Comparison of the KL divergence under the MLE $\mymletheta$, $L_2$ regularization and ICE regularization $\myregtheta_{ICE}$ against a large sample size for the case when $p=10$ and $m=4$. }}
\label{tab:converge}
\end{table}

Generally the $ \myregtheta_{ICE} $ estimates are seen to converge slightly faster than the $ \mymletheta $ estimates. The regularization in $ \myregtheta_{L_{2}} $ is observed to be beneficial for very small sample sizes, but then becomes marginally detrimental for large $ n $.


\section{Optimized Computation Results}
\label{sect:optObj}

For any model satisfying White's Regularity Criteria, it is known that the matrix $ J $ is positive definite near the MLE optimum $ \mymletheta $. This implies that $ J $ is diagonally dominated, and indeed considering just its diagonal elements $ D $, it is known that $ tr(ID^{-1}) > 0 $. Indeed $ tr(ID^{-1}) $ differs strongly from $ tr(IJ^{-1}) $ most strongly for models with strong regressor interactions. Therefore, using finite difference gradients, consider the following approximations for the ICE objective function:

\begin{enumerate}
\item $ \myregtheta $: $ J $ is computed directly, ICE is implemented as written. 
\item $ \myregtheta_{2} $: $ J $ is taken to be constant w.r.t. $ \mytheta $: ($ J_{\mytheta} = J_{\mymletheta}$).
\item $ \myregtheta_{3} $: $ J $ is taken to be diagonal: ($ J = D $).
\item $ \myregtheta_{4} $: $ J $ is taken to be the identity: ($ J = \mathcal{I} $).
\end{enumerate}

Clearly, we expect that $ \myregtheta_{4} $ above is the least accurate approximation, and items $ \myregtheta_{2} $ and $ \myregtheta_{3} $ have varying levels of accuracy depending on the problem at hand. The cost comparison of these approaches is shown in Table \ref{tab:approxCost}. 

\begin{table}[H]
\begin{tiny}
\begin{tabular}{cccccc}
\hline
\textbf{Approximation} & \textbf{Objective Cost (Space)} & \textbf{Objective Cost (Time)} & \textbf{Gradient Cost (Space) }& \textbf{Gradient Cost (Time)} \\
\midrule
Direct Computation & $ O(p^{2}) $ & $ O(p^{3}) $  & $ O(p^{2}) $ & $ O(p^{4}) $ \\
$ J_{\mytheta} = J_{\mymletheta} $ & $ O(p^{2}) $ & $ O(p^{2}) $  & $ O(p^{2}) $ & $ O(p^{3}) $ \\
$ J = D $ & $ O(p) $ & $ O(p) $  & $ O(p) $ & $ O(p^{2}) $ \\
$ J = \mathcal{I} $ & $ O(p) $ & $ O(p) $ & $ O(p) $ & $ O(p^{2}) $ \\
\hline
\end{tabular}
\caption{The asymptotic computational cost (per iteration) of various proposed approximations as a function of parameter count $p$. Cost is amortized when $J_{\mytheta} = J_{\mymletheta}$ assuming that $ n \approx p $. Note that a typical model will cost $ O(p) $ in time and space for both the objective function and its gradients.}
\label{tab:approxCost}
\end{tiny}
\end{table}

\begin{remark}
When computing gradients for use in a solver, often approximation error will have only a marginal impact on the final result, though it may increase the number of iterations needed for convergence. Broyden's method \cite{Broyden65} is a typical example of this approach in action. Efficient approximations of $  [\partial_{\mytheta} \hat{J}] $ might similarly have only a minor effect on accuracy and iteration count. The construction of approximate analytical derivatives is beyond the scope of the present work. 
\end{remark}

These approximations were computed and compared for the Friedman (see \mbox{Section~\ref{FriedmanTest}}) problem, and the results are shown in Table \ref{tab:approxCompare} below.

\begin{table}[H]
\begin{tiny}
\begin{tabular}{cccccc}
\hline
\textbf{\emph{n}} & \boldmath{$ \rho_{KL}(f, g_{\mymletheta}) $}  & \boldmath{$ \rho_{KL}(f, g_{\myregtheta}) $}  & \boldmath{$ \rho_{KL}(f, g_{\myregtheta_{2}}) $}  & \boldmath{$ \rho_{KL}(f, g_{\myregtheta_{3}}) $}  & \boldmath{$ \rho_{KL}(f, g_{\myregtheta_{4}}) $}   \\
\midrule
$ 8 $ & $ 1.22 \times 10^{+1} $ & $ 4.55 \times 10^{+0} ($-$4.89) $ & $ 5.70 \times 10^{+0} ( -5.26 ) $ & $ 3.83 \times 10^{+0} ( -5.22 ) $ & $ 1.28 \times 10^{+0} ( -4.67 ) $ \\
$ 16 $ & $ 6.68 \times 10^{-1}  $ & $ 2.99 \times 10^{-1}  ( -8.13 )  $ & $ 3.53 \times 10^{-1}  ( -10.56 )  $ & $ 3.36 \times 10^{-1}  ( -8.30 )  $ & $ 5.47 \times 10^{-1}  ( -2.11 )  $ \\
$ 32 $ & $ 1.45 \times 10^{-1}  $ & $ 1.14 \times 10^{-1}  ( -6.90 )  $ & $ 1.04 \times 10^{-1}  ( -8.18 )  $ & $ 1.08 \times 10^{-1}  ( -10.16 )  $ & $ 3.63 \times 10^{-1}  ( 19.60 )  $  \\
$ 64 $ & $ 5.93 \times 10^{-2}  $ & $ 4.80 \times 10^{-2}  ( -10.42 )  $ & $ 4.70 \times 10^{-2}  ( -6.95 )  $ & $ 4.81 \times 10^{-2}  ( -9.81 )  $ & $ 2.38 \times 10^{-1}  ( 39.08 )  $ \\
$ 128 $ & $ 2.48 \times 10^{-2}  $ & $ 2.24 \times 10^{-2}  ( -6.38 )  $ & $ 2.33 \times 10^{-2}  ( -2.26 )  $ & $ 2.26 \times 10^{-2}  ( -6.00 )  $ & $ 1.62 \times 10^{-1}  ( 61.85 )  $ \\
$ 256 $ & $ 1.21 \times 10^{-2}  $ & $ 1.16 \times 10^{-2}  ( -4.28 )  $ & $ 1.20 \times 10^{-2}  ( -0.68 )  $ & $ 1.16 \times 10^{-2}  ( -4.11 )  $ & $ 1.01 \times 10^{-1}  ( 68.82 )  $ \\
$ 512 $ & $ 6.26 \times 10^{-3}  $ & $ 6.10 \times 10^{-3}  ( -2.41 )  $ & $ 6.16 \times 10^{-3}  ( -0.84 )  $ & $ 6.10 \times 10^{-3}  ( -2.39 )  $ & $ 5.42 \times 10^{-2}  ( 63.84 )  $ \\
$ 1024 $ & $ 3.05 \times 10^{-3}  $ & $ 3.00 \times 10^{-3}  ( -2.66 )  $ & $ 3.04 \times 10^{-3}  ( -0.37 )  $ & $ 2.99 \times 10^{-3}  ( -2.73 )  $ & $ 2.61 \times 10^{-2}  ( 59.78 )  $ \\
\hline
\end{tabular}
\caption{Comparison of the average KL divergence across $ 200 $ replications for MLE and several variants of ICE given a fitting set size of $ n $. For estimators other than $ \mymletheta $, the values in parentheses denotes the t-statistic of the difference between this estimator and $ \mymletheta $, with negative values indicating that the listed estimator has a lower KL divergence.}
\label{tab:approxCompare}
\end{tiny}
\end{table}

From Table \ref{tab:approxCompare}, it is apparent that approach $ \myregtheta_{4} $, taking $ J = \mathcal{I} $ is not effective. This is not surprising as the actual $ J $ matrix has dramatic differences in scale between regressors. Approximation $ \myregtheta_{3} $, taking $ J = D $ is accurate enough that it cannot be statistically distinguished from the direct computation of ICE by the test above. Approximation $ \myregtheta_{2} $ tends to underperform approximation (3). 

Therefore, we propose taking $ J = D $ as a more numerically stable approximation of the ICE objective.

\section{Conclusions} \label{sect:conclusion}

Takeuchi \cite{Takeuchi1976} is believed to be the first to have proposed using an objective function similar to ICE in order to reduce generalization error, though it was applied via model selection. Firth \cite{Firth93} introduced a similar term to reduce parameter bias in model fitting, as opposed to model selection, though he derived it only for exponential model families and did not consider its effect on generalization error. It is not known why this approach did not find widespread use, but one may infer that the $ O(p^{4}) $ computational cost and instability was enough to keep it from wider adoption. 

In this paper, we reintroduce the objective function of \cite{Takeuchi1976} and provide a more general proof of its widespread applicability. We then show that efficient implementations costing only $ O(p) $ are possible. Under finite sample sizes, this bias correction term is shown experimentally in several models to lead to significant reduction in bias compared to maximum likelihood estimation with and without $L_2$ regularization. ICE offers many advantages over $ L_{2} $ penalized maximum likelihood estimation: (i) it's suitable for most nonlinear models, (ii) it's provably asymptotically convergent; and (iii) does not rely on any parameters which would need to be provided by the operator or deduced through cross-validation. 

\section*{Acknowledgements}

The authors would like to thank Bertrand Nortier for his suggestions, feedback, and assistance. We would also like to thank Kevin Atteson for providing early feedback that steered us away from error.

\vspace{6pt} 



\appendix

\section{Proofs}
\label{sect:supp}


\subsection{White's Regularity Conditions}
\label{sect:white} 

\begin{defn}[White's regularity conditions]
White \cite{White1982} provides the following regularity conditions:
\begin{itemize}
\item[A1:] The independent random vectors, $X_i, i= 1,\dots,n$, have common joint distribution function $F$ on $\Omega$, a  measurable Euclidean space, with measurable Radon--Nikodym density $f = dF/dx$.
\item[A2:]  \textls[-15]{The family of distribution functions $G(x | \mytheta)$ has Radon--Nikodym densities $g(x | \mytheta) = dG(x | \mytheta)/dx$} which are measurable in x  for every $\mytheta \in \Theta$, a  compact subset of p-dimensional Euclidean space, and continuous in $\mytheta$ for every $x\in\Omega$.

\item[A3:] (a) $\myE[log~ f(X)]$ exists and $| log~g(x | \mytheta)| \leq m(x),~ \forall \mytheta \in \Theta$, where m is integrable with respect to F; (b) $\rho_{KL}(f, g_{\mytheta}) $ has a  unique minimum at $\mytheta_0 \in \Theta$.

\item[A4:] $\partial_{\mytheta} (log~ g(x | \mytheta))$ are measurable functions of $x$ for each $\mytheta \in \Theta$ and continuously differentiable functions of $\mytheta$  for each $x\in\Omega$.

\item[A5:] $|\partial^2_{\mytheta} (log ~g(x | \mytheta))|$ and  $|\partial_{\mytheta}(log~ g(x | \mytheta)) \cdot \partial_{\mytheta} (log~ g(x | \mytheta) |)$  are dominated by functions integrable in $x$
 with respect to F for all $x\in\Omega$ and $\mytheta \in \Theta$.
\item[A6:] (a) $\mytheta_0$ is an interior point of the parameter space; (b) $\myE[\partial_{\mytheta} (log~ g(x | \mytheta)) \cdot \partial_{\mytheta} (log~ g(x | \mytheta))]$ is non-singular; (c) $\theta_0$ is a regular point of $\myE[\partial^2_{\mytheta} (log~ g(x | \mytheta))]$.
\end{itemize}
\end{defn}


\subsection{Proof of Finite Variance}
\label{sect:fvproof}

\begin{lemma}[Finite variance]

Suppose the following conditions hold:

\begin{enumerate}
\item $\mathcal{M}$ satisfies White's regularity conditions A1--A6 (see Section \ref{sect:white} or \cite{White1982}). 
\item $ \mytheta_0 $ is a global minimum of $ -\mathcal{L}(\mytheta) $ in the compact space $ \Theta $ defined in A2. 
\item There exists a $ \varepsilon > 0 $ such that $ -\mathcal{L}(\mytheta_0) < -\mathcal{L}(\mytheta_1) - \varepsilon $ for all other local minima $ \mytheta_1 $.
\item For $ k = {0, 1, 2, 3, 4, 5} $ the derivative $ \partial^{k}_{\mytheta} \mathcal{L}(\mytheta) $ exists, is continuous, and bounded on an open set around $ \mytheta_0 $. 
\item For $ k = {0, 1, 2, 3, 4, 5} $, the variance $ \myV[\partial^{k}_{\mytheta} \ell(\mytheta, \Var_n)] \rightarrow 0 $ as $ n \rightarrow \infty $ on an open set around $ \mytheta_0 $. 
\end{enumerate}

Then, for sufficiently large $ n $ there exists a compact subset $ U \subset \Theta $ containing $ \mytheta_0, \mymletheta, $ such that

\begin{enumerate}
\item For $ k = {0, 1, 2, 3} $ the derivative $ \partial^{k}_{\mytheta} \ell^{*}(\mytheta, \data) $ exists, is continuous, and bounded on $ U $, almost surely. 
\item For $ k = {0, 1, 2, 3} $, $ \myV[\partial^{k}_{\mytheta} \ell^{*}(\mytheta, \Var_n)] \rightarrow 0 $ as $ n \rightarrow \infty $ on $ U $, almost surely. 
\item $ \myregtheta \in U $ as $ n \rightarrow \infty $ almost surely. 
\end{enumerate}

\label{asymptoticLemma}
\end{lemma}
 
\begin{proof}

Assumptions (4) and (5) establish existence of some set, $ S $, containing $ \mytheta_0 $ such that $ \mathcal{L}(\mytheta) $ is bounded on $ S $, and its estimate, $ \ell(\mytheta, \Var_n) $ has finite variance. Therefore, $ -\ell(\mytheta, \data) $ is also bounded on $ S $ almost surely. Similarly for the first 5 derivatives. White's criteria imply that $ J_{\mytheta_0} $ is positive definite on an open set around $ \mytheta_0 $, and thus one can form a compact set $ U \subset S $ containing an open set around $ \mytheta_0 $ on which the minimum eigenvalue of $ J_{\mytheta} $ is bounded away from 0.

Note that $ J_{\mytheta} $ is three times differentiable on $ U $ by Assumption 4, as is $ \hat{J}_{\mytheta} $, as established above. Then $ \hat{J}^{-1}_{\mytheta} $ is also positive definite and bounded on $ U $. It can be shown to also have three derivatives by using the well-known matrix relation

\begin{equation}
\partial_{\mytheta} A^{-1} = -A^{-1} (\partial_{\mytheta} A) A^{-1}.
\end{equation}

It follows that $ \hat{J}^{-1}_{\mytheta} $ is also positive definite, nonsingular, and bounded on $ U $. Similarly for $ \hat{I}_{\mytheta} $, and thus $ tr(\hat{I}_{\mytheta}\hat{J}^{-1}_{\mytheta}) $ is bounded with finite variance on $ U $.  It also has three bounded derivatives with finite variance.

Therefore, $ -\ell^{*}(\mytheta, \Var_n) \rightarrow -\ell(\mytheta, \Var_n) $, and $ \myregtheta \rightarrow \mytheta_0 $ as $ n \rightarrow \infty $, with the convergence being in probability. This means that $ U $ contains $ \mytheta_0 $, $ \mymletheta $, and $ \myregtheta $ almost surely for large enough $ n $. Similarly, on $ U $ we have three continuous, bounded derivatives of $ \partial^{k}_{\mytheta} \ell^{*}(\mytheta, \data)  $ almost surely. 
\end{proof}


\subsection{Proof of Asymptotic Normality}
\label{sect:anproof}

\begin{theorem}[Asymptotic Normality]
Provided the conditions hold in Lemma \ref{asymptoticLemma}, namely, that 

\begin{enumerate}
\item For $ k = {0, 1, 2, 3} $ the derivative $ \partial^{k}_{\mytheta} \ell^{*}(\mytheta, \data) $ exists, is continuous, and bounded on $ U $. 
\item For $ k = {0, 1, 2, 3} $, $ \myV[\partial^{k}_{\mytheta} \ell^{*}(\mytheta, \data)] \rightarrow 0 $ as $ n \rightarrow \infty $ on $ U $. 
\end{enumerate}

Then 

$$ \sqrt{n}(\myregtheta -\myregtheta_0)  \rightarrow N(0,  (\hat{J}_{\myregtheta_{0}}^{*})^{-1}\hat{I}^{*}_{\myregtheta_{0}}(\hat{J}_{\myregtheta_{0}}^{*})^{-1}).$$

\label{normalityTheorem}
\end{theorem}
\begin{proof}

As the first derivatives of $ \ell $ are continuous, the mean value theorem may be applied: 

\be
\partial_{\mytheta} \ell^{*}(\myregtheta_0) = \partial_{\mytheta} \ell^{*}(\myregtheta) + (\myregtheta - \myregtheta_0) \hat{J}^{*}_{\bar{\mytheta}} = (\myregtheta - \myregtheta_0) \hat{J}^{*}_{\bar{\mytheta}}.
\ee

$ \bar{\mytheta} $ is between $ \myregtheta $ and $ \myregtheta_0 $. Under the assumptions of Lemma \ref{asymptoticLemma}, and given its finite variance, $ \hat{J}_{\bar{\mytheta}} $ is almost surely (in the large $ n $ limit) positive definite, and thus invertible as both $ \myregtheta $ and $ \myregtheta_0 $ are in $ U $, and $ \bar{\mytheta} $ is between them. Therefore, 

\be
 (\myregtheta - \myregtheta_0) = (\hat{J}^{*}_{\bar{\mytheta}})^{-1} \partial_{\mytheta} \ell^{*}(\myregtheta_0).
\ee

Applying the mean value theorem a second time gives

\be
\hat{J}_{\bar{\mytheta}} = \hat{J}^{*}_{\myregtheta_0} +  (\bar{\mytheta} - \myregtheta_0) \hat{J}^{*}_{\mytheta_1},
\ee
with $ \mytheta_1 $ between $ \bar{\mytheta} $ and $ \myregtheta_0 $. If the order of $ (\myregtheta - \myregtheta_0) = O_{p}(\delta) $, where $\delta:=n^{-1/2}$, then 

\be
\hat{J}^{*}_{\bar{\mytheta}} = \hat{J}^{*}_{\myregtheta_0} +  O_{p}(\delta).
\ee

As all of the $ \hat{J}^{*} $ are bounded away from zero in probability, we have

\be
(\hat{J}^{*}_{\bar{\mytheta}})^{-1} = (\hat{J}^{*}_{\myregtheta_0})^{-1} +  O_{p}(\delta),
\ee
with the equality holding in probability. In the large $ n $ limit, $ \delta \rightarrow 0 $, and thus

\be
 (\myregtheta - \myregtheta_0) = (\hat{J}^{*}_{\myregtheta_0})^{-1} \partial_{\mytheta} \ell^{*}(\myregtheta_0).
 \label{normalPrecursor}
\ee

As $ \partial_{\mytheta} \ell^{*}(\myregtheta_0) $ is the sum of $ n $ independent vectors, it is asymptotically normally distributed by the central limit theorem, and its mean is $ 0 $ by the definition of $ \myregtheta_0 $. Its variance is therefore $ \myV[\partial_{\myregtheta} \ell^{*}(\myregtheta_0)] = \myE[\partial_{\mytheta} \ell^{*}(\myregtheta_0) (\partial_{\mytheta} \ell^{*}(\myregtheta_0))^{T}] = \frac{1}{n} \hat{I}_{\myregtheta_0} $

Substituting this into Equation (\ref{normalPrecursor}) yields

\be
 \sqrt{n}(\myregtheta - \myregtheta_0) = N(0, (\hat{J}^{*}_{\myregtheta_0})^{-1} \hat{I}^{*}_{\myregtheta_0} (\hat{J}^{*}_{\myregtheta_0})^{-1}),
\ee
establishing the result. 
\end{proof}


\subsection{Proof of Prediction Bias Order under ICE}
\label{sect:pboproof}

\begin{theorem}[Prediction Bias Estimation under ICE]
\label{biasTheorem}
By minimizing $ -\ell^{*} $ instead of $ -\ell $, the first order terms of the prediction bias are cancelled leaving a $ O_{p}(n^{-3/2}) $ residual term
\be
-\mathcal{L}(\hat{\myregtheta}) =  -\ell(\myregtheta(\Var_{n}), \Var_{n}) + O_{p}(n^{-3/2}).
\ee
\end{theorem}

\begin{proof}
Note that in Takeuchi's proof \cite{Takeuchi1976}, the use of $ \mymletheta $ was prescribed. Therefore, this approach could be used only for model selection, and not for model fitting. Now consider the bias under the ICE estimator $ \myregtheta $. 
\begin{eqnarray*}
b(\myregtheta(\bVar_n), \bVar_n) &=&
\myE_{\bVar_n}\left[\log \model{\bVar_n,\myregtheta(\bVar_n)}-\log \model{\bVar_n,\mytheta_0}\right]\\
&+&\myE_{\bVar_n}\left[\log \model{\bVar_n,\mytheta_0} -n\myE_{\newVar_n}[\log \model{\newVar_n,\mytheta_0}]\right]\\
&+&n\myE_{\bVar_n}\left[\myE_{\newVar_n}[\log \model{\newVar_n,\mytheta_0}] -\myE_{\newVar_n}[\log \model{\newVar_n,\myregtheta(\bVar_n)}]\right].
\label{bias}
\end{eqnarray*}

As the second term is zero, this can be simplified to 
\begin{eqnarray*}
b(\myregtheta(\bVar_n), \bVar_n) &=&  -n \myE_{\bVar_n}\left[\ell(\myregtheta(\bVar_n), \Var_n) - \ell(\mytheta_0, \Var_n)\right] \\
&-& n \myE_{\bVar_n}\left[\mathcal{L}(\mytheta_0) -\mathcal{L}(\myregtheta(\bVar_n))\right].
\label{bias}
\end{eqnarray*}

Define $ \delta = \frac{1}{\sqrt{n}} $, and then recall from White \cite{White1982} that $ (\mymletheta-\mytheta_0) $ is $O_{p}(\delta) $. Similarly, recall from Theorem \ref{normalityTheorem} that $  (\myregtheta-\myregtheta_0) $ is also $ O_{p}(\delta) $. As constructed, the error term $ b(\myregtheta(\bVar_n), \bVar_n) $ is $ O_{p}(1) $ (actually $ O(1) $ as it is an expectation), and terms of order $ O_{p}(\delta) $ and higher will be dropped. Therefore, as in Takeuchi's derivation \cite{Takeuchi1976}, the Taylor expansions below will be truncated at second order in $ \delta $, dropping terms of order $ O_{p}(\delta^{3}) $ or higher. Additionally, we will occasionally drop indications of $ \bVar_n $ where the meaning is clear and it greatly simplifies the notation. 

With that truncation, recalling that $ \mytheta_0 $ is a minimum of $ \mathcal{L}(\mytheta) $ and thus has zero gradient: 
\begin{equation}
n \myE_{\bVar_n}\left[\mathcal{L}(\mytheta_0) -\mathcal{L}(\myregtheta(\bVar_n))\right] = \frac{n}{2}\myE_{\bVar_n}[(\myregtheta-\mytheta_0)^TJ_{\mytheta_0}(\myregtheta-\mytheta_0) ] + O(\delta).
\end{equation}

Recall Theorem \ref{normalityTheorem}, and recall the that for the quadratic form: 

\begin{equation}
\myE[\varepsilon^{T}A\varepsilon] = tr(A\Sigma) + \mu^{T}A\mu, \qquad \myE[\varepsilon] = \mu,  \myV[\varepsilon] = \Sigma.
\label{eqn:quad_form}
\end{equation}

Therefore,
\begin{eqnarray*}
\frac{n}{2}\myE_{\bVar_n}[(\myregtheta-\mytheta_0)^TJ_{\mytheta_0}(\myregtheta-\mytheta_0) ] 
&=& \frac{1}{2} tr(J_{\mytheta_0}[(J_{\myregtheta_{0}}^{*})^{-1}I^{*}_{\myregtheta_{0}}(J_{\myregtheta_{0}}^{*})^{-1}) ]) \\
&+& \frac{n}{2} (\myregtheta_0-\mytheta_0)^TJ_{\mytheta_0}(\myregtheta_0-\mytheta_0) \\
&+& O(\delta).
\end{eqnarray*}

Now note that the second term on the right is a constant, and therefore would take no part in any optimization. Therefore, it can be safely ignored and

\be
n \myE_{\bVar_n}\left[\mathcal{L}(\mytheta_0) -\mathcal{L}(\myregtheta(\bVar_n))\right]
= \frac{1}{2} tr(J_{\mytheta_0}[(J_{\myregtheta_{0}}^{*})^{-1}I^{*}_{\myregtheta_{0}}(J_{\myregtheta_{0}}^{*})^{-1}) ]) + O(\delta).
\ee

Addressing the first term, again taking a Taylor expansion, we find that 
\be
\ell(\mytheta_0) = \ell(\myregtheta)  +  (\mytheta_0-\myregtheta)^T \partial_{\mytheta}\ell(\myregtheta) + \frac{1}{2} (\mytheta_0-\myregtheta)^T\partial^2_{\mytheta}\ell(\myregtheta)(\mytheta_0-\myregtheta)  + O_{p}(\delta^{3}).
\ee

Therefore, recalling that $ n O_{p}(\delta^{3}) = O_{p}(\delta) $ gives

\begin{eqnarray*}
n \myE_{\bVar_n}\left[\ell(\myregtheta(\bVar_n), \Var_n) - \ell(\mytheta_0, \Var_n)\right] 
 &=&  \frac{1}{2} tr(J_{\myregtheta}[(J_{\myregtheta_0}^{*})^{-1}I^{*}_{\myregtheta_0}(J_{\myregtheta_0}^{*})^{-1}) ]) \\
 &+& \frac{n}{2} (\myregtheta_0-\mytheta_0)^TJ_{\myregtheta}(\myregtheta_0-\mytheta_0) \\
 &+& n \myE_{\bVar_n}[(\myregtheta - \mytheta_0)^T \partial_{\mytheta}\ell(\myregtheta)] \\
 &+& O_{p}(\delta).
\end{eqnarray*}

Now examine the last term:
\begin{eqnarray*}
n \myE_{\bVar_n}[(\myregtheta - \mytheta_0)^T \partial_{\mytheta}\ell(\myregtheta)] 
&=&  n \myE_{\bVar_n}[(\myregtheta - \myregtheta_0)^T \partial_{\mytheta}\ell(\myregtheta)] \\
&+& n \myE_{\bVar_n}[(\myregtheta_0 - \mytheta_0)^T \partial_{\mytheta}\ell(\myregtheta)].
\end{eqnarray*}

However, $ (\myregtheta_0 - \mytheta_0) $ does not depend on $ \bVar_n $, so it can be pulled out of the expectation, then substitute in a first order Taylor expansion
\begin{eqnarray*}
\myE_{\bVar_n}[(\myregtheta_0 - \mytheta_0)^T \partial_{\mytheta}\ell(\myregtheta)] 
&=& (\myregtheta_0 - \mytheta_0)^T \myE_{\bVar_n}[ \partial_{\mytheta}\ell(\myregtheta)] \\
&=& (\myregtheta_0 - \mytheta_0)^T \myE_{\bVar_n}[ \partial_{\mytheta}\ell(\myregtheta_{0}) + (\myregtheta - \myregtheta_0)\partial^{2}_{\mytheta}\ell(\myregtheta_{0}) + O_{p}(\delta^{2})] \\
&=&  (\myregtheta_0 - \mytheta_0)^T  \partial_{\mytheta}\mathcal{L}(\myregtheta_{0}) \\
&+& (\myregtheta_0 - \mytheta_0)^T \myE_{\bVar_n}[ (\myregtheta - \myregtheta_0)] \partial^{2}_{\mytheta} \mathcal{L}(\myregtheta_{0}) + O(\delta^{3}) \\
&=&  (\myregtheta_0 - \mytheta_0)^T  \partial_{\mytheta} \mathcal{L}(\myregtheta_{0})  + O(\delta^{3}),
\end{eqnarray*}

with the last equality following from the fact that $ \myE_{\bVar_n}[ (\myregtheta - \myregtheta_0)] = 0 $ and the substitution of $  (\myregtheta_0 - \mytheta_0)^T O(\delta^{2}) = O(\delta^{3}) $. This term is therefore a constant, up to $ O(\delta^{3}) $, and takes no part in optimization of $ \mytheta $, thus it can be dropped from further consideration. Therefore $ n \myE_{\bVar_n}[(\myregtheta_0 - \mytheta_0)^T \partial_{\mytheta}\ell(\myregtheta)]  = O(\delta) $, and 
\begin{eqnarray*}
n \myE_{\bVar_n}[(\myregtheta - \mytheta_0)^T \partial_{\mytheta}\ell(\myregtheta)] 
&=&  n \myE_{\bVar_n}[(\myregtheta - \myregtheta_0)^T \partial_{\mytheta}\ell(\myregtheta)] \\
&+& O(\delta).
\end{eqnarray*}

Recombining these terms yields
\begin{eqnarray*}
n \myE_{\bVar_n}\left[\ell(\myregtheta(\bVar_n), \Var_n) - \ell(\mytheta_0, \Var_n)\right] 
 &=&  \frac{1}{2} tr(J_{\myregtheta}[(J_{\myregtheta}^{*})^{-1}I^{*}_{\myregtheta}(J_{\myregtheta}^{*})^{-1}) ]) \\
 &+& \frac{n}{2} (\myregtheta_0-\mytheta_0)^TJ_{\myregtheta}(\myregtheta_0-\mytheta_0) \\
 &+&  n \myE_{\bVar_n}[(\myregtheta - \myregtheta_0)^T \partial_{\mytheta}\ell(\myregtheta)] \\
 &+& O(\delta).
\end{eqnarray*}

Thus the bias (neglecting the constant terms) is then
\begin{eqnarray*}
b(\myregtheta(\bVar_n), \bVar_n) 
 &=& -\frac{1}{2} tr(J_{\myregtheta}[(J_{\myregtheta}^{*})^{-1}I^{*}_{\myregtheta}(J_{\myregtheta}^{*})^{-1}) ]) \\
 &-& \frac{1}{2} tr(J_{\mytheta_0}[(J_{\myregtheta_{0}}^{*})^{-1}I^{*}_{\myregtheta_{0}}(J_{\myregtheta_{0}}^{*})^{-1}) ]) \\
 &-& \frac{n}{2} (\myregtheta_0-\mytheta_0)^TJ_{\myregtheta}(\myregtheta_0-\mytheta_0) \\
 &-& n \myE_{\bVar_n}[(\myregtheta - \myregtheta_0)^T \partial_{\mytheta}\ell(\myregtheta)] \\
 &+& O(\delta).
\label{bias}
\end{eqnarray*}

Because $ \ell^{*}(\mytheta) = \ell(\mytheta) + O_{p}(\delta^{2}) $, it follows that $ J_{\mytheta}^{*} = J_{\mytheta} + O_{p}(\delta^{2}) $ and thus $ J_{\myregtheta}(J_{\myregtheta}^{*})^{-1} = I + O(\delta^{2}) $. Similarly for $ J_{\mytheta_0} $. In addition $ I_{\myregtheta} = I_{\myregtheta_0} + O(\delta) $,  thus the two trace terms can be simplified and combined: 

\begin{eqnarray*}
b(\myregtheta(\bVar_n), \bVar_n) 
 &=& - tr(I_{\myregtheta}J_{\myregtheta}^{-1}) \\
 &-& \frac{n}{2} (\myregtheta_0-\mytheta_0)^TJ_{\myregtheta}(\myregtheta_0-\mytheta_0) \\
 &-& n \myE_{\bVar_n}[(\myregtheta - \myregtheta_0)^T \partial_{\mytheta}\ell(\myregtheta)] \\
 &+& O(\delta).
\label{bias}
\end{eqnarray*}

As $ \myregtheta_0 $ is a minimum of $ \mathcal{L}^{*} $, begin by Taylor expanding the derivatives

\be
\partial_{\mytheta} \mathcal{L}(\myregtheta_{0}) = \partial_{\mytheta} \mathcal{L}(\mytheta_{0}) - (\myregtheta_{0} - \mytheta_0) J_{\mytheta_0} = - (\myregtheta_{0} - \mytheta_0) J_{\mytheta_0}.
\ee

Moreover, show from the definition of $  \mathcal{L}^{*} $ that 

\be
\partial_{\mytheta} \mathcal{L}^{*}(\myregtheta_{0}) = 0 = \partial_{\mytheta} \mathcal{L}(\myregtheta_{0}) + \frac{1}{n} \partial_{\mytheta} tr(IJ^{-1})
\ee

Then, after Taylor expanding $ \mathcal{L}(\myregtheta_{0}) $ around $ \mytheta_{0} $, it is seen that

\be
(\myregtheta_{0} - \mytheta_0)  = \frac{1}{n} J^{-1}_{\mytheta_0} \partial_{\mytheta} tr(IJ^{-1}).
\label{meanDiffEquation}
\ee

Noting that $ J^{-1}_{\mytheta} = O(1) $ and $ tr(IJ^{-1}) = O(1) $, it holds that $ (\myregtheta_{0} - \mytheta_0) = O(\delta^{2}) $. 

Therefore, $  \frac{n}{2} (\myregtheta_0-\mytheta_0)^TJ_{\myregtheta}(\myregtheta_0-\mytheta_0)  = O(\delta^{2}) $, and can be neglected. 

Then, the bias becomes
\begin{eqnarray*}
b(\myregtheta(\bVar_n), \bVar_n) 
 &=& - tr(I_{\myregtheta}J_{\myregtheta}^{-1}) \\
 &-& n \myE_{\bVar_n}[(\myregtheta - \myregtheta_0)^T \partial_{\mytheta}\ell(\myregtheta)] \\
 &+& O(\delta).
\label{bias}
\end{eqnarray*}

However, for the same reason, $ \partial_{\mytheta}\ell(\myregtheta) = O(\delta^{2}) $, so the last term 
$  n \myE_{\bVar_n}[(\myregtheta - \myregtheta_0)^T $ $\partial_{\mytheta}\ell(\myregtheta)] = O(\delta) $, and it too can be absorbed into the residual. 

Therefore, 
\begin{eqnarray*}
b(\myregtheta(\bVar_n), \bVar_n) 
 &=& - tr(I_{\myregtheta}J_{\myregtheta}^{-1}) \\
 &+& O(\delta),
\label{bias}
\end{eqnarray*}
and
\be
-\mathcal{L}(\hat{\mytheta}) =  -\ell(\myregtheta(\newVar_{n}), \newVar_{n}) + \frac{1}{n}  tr(I_{\hat{\mytheta}}J_{\hat{\mytheta}}^{-1}) + O_{p}(\delta^{3}) + C,
\ee
where the constant $ C $ is composed of the neglected constant terms from earlier stages
\begin{eqnarray*}
C
 &=&  -\frac{1}{2} (\myregtheta_0-\mytheta_0)^TJ_{\mytheta_0}(\myregtheta_0-\mytheta_0) \\
 &+& -(\myregtheta_0 - \mytheta_0)^T  \partial_{\mytheta} \mathcal{L}(\myregtheta_{0}).
\label{bias}
\end{eqnarray*}

However, the last term is $O(\delta^{3}) $, and the first is $ O(\delta^{2}) $, so this may be approximated as 

\be
C = -\frac{1}{2} (\myregtheta_0-\mytheta_0)^TJ_{\mytheta_0}(\myregtheta_0-\mytheta_0).
\ee

Recalling again that $ (\myregtheta_0-\mytheta_0) = O(\delta^{2}) $, it is clear that $ C = O(\delta^{4}) $, and may thus be absorbed into the $ O(\delta^{3}) $ residual term. Therefore, 

\be
-\mathcal{L}(\hat{\mytheta}) =  -\ell(\myregtheta(\newVar_{n}), \newVar_{n}) + \frac{1}{n}  tr(I_{\hat{\mytheta}}J_{\hat{\mytheta}}^{-1}) + O(\delta^{3}).
\ee

Comparing this to the form of $ \ell^{*}(\mytheta) $:

\be
-\mathcal{L}(\hat{\mytheta}) =  -\ell^{*}(\myregtheta(\newVar_{n}), \newVar_{n}) + O_{p}(\delta^{3}).
\ee

Thus, by minimizing $ -\ell^{*} $ instead of $ -\ell $, the first order terms of the prediction bias are canceled and, in expectation, a more accurate model is produced. 
\end{proof}






\end{document}